\documentclass[12pt]{article}
\usepackage{a4wide}
\usepackage{cite}
\usepackage{amsmath}
\usepackage{amstext,amsfonts,amsbsy,eucal,amssymb}
\usepackage{amsmath,pifont}
\usepackage{color}
\usepackage[footnotesize,hang,sf]{caption}
\usepackage{xypic}
\xyoption{all}
\usepackage{epsfig,psfrag}

\usepackage{
times,
txfonts
}
\usepackage{youngtab}
\numberwithin{equation}{section}
\renewcommand{\thefootnote}{\fnsymbol{footnote}}

\def\openone{\leavevmode\hbox{\small1\kern-3.8pt\normalsize1}}%

\DeclareMathOperator{\wt}{wt}
\DeclareMathOperator{\ex}{ex}
\DeclareMathOperator{\ps}{ps}
\DeclareMathOperator{\length}{\ell}
\DeclareMathOperator{\Prob}{Prob}

\begin{document}

\baselineskip 21pt
\parskip 7pt

\hfill revised on November 23, 2002

\vspace{24pt}

\begin{center}

  {\Large\textbf{
      Vicious Walkers and Hook Young Tableaux
      }
    }

  \vspace{24pt}

  {\large Kazuhiro \textsc{Hikami}
  \footnote[2]{
    \texttt{hikami@phys.s.u-tokyo.ac.jp}
    }
    and
    Takashi \textsc{Imamura}
  \footnote[3]{
    \texttt{imamura@monet.phys.s.u-tokyo.ac.jp}
    }
  }

  \vspace{2pt}
   \textsl{Department of Physics, Graduate School of Science,\\
     University of Tokyo,\\
     Hongo 7--3--1, Bunkyo, Tokyo 113--0033, Japan.
     }

(Received: September 20, 2002)

\vspace{18pt}

\noindent
\underline{\textsf{ABSTRACT}}
\end{center}

We consider a generalization of the vicious walker model.
Using a bijection map between the path configuration of the
non-intersecting random walkers and the hook Young diagram,
we compute the probability concerning the number of walker's movements.
Applying the saddle point method, we reveal that the scaling limit
gives the Tracy--Widom distribution, which is same with the limit
distribution of the largest
eigenvalues of the Gaussian unitary ensemble.

\vfill
\noindent

\noindent
\textsf{PACS:}

\newpage

\renewcommand{\thefootnote}{\arabic{footnote}}

\section{Introduction}

Since it was shown that the path configuration of the random vicious
walkers~\cite{MEFish84a} is related with the Young
tableaux~\cite{GuttOwczVien98a,PJForre01a,JBaik00a},
much attention has been paid on the statistical combinatorial problems
which
are  intimately related with the Young tableaux.
Among them is
the random permutation~\cite{BaikDeifJoha99a},
the random word~\cite{TracWido99b},
the point process~\cite{KJohan99a,Seppa01a},
the random growth model
(the polynuclear growth model, oriented digital boiling
model)~\cite{PrahoSpohn00a,GravTracWido00a},
the queuing theory~\cite{Barys01a},
and so on.
Interesting is that the scaling limits of these models have the
universality
that the fluctuation is of order $N^{1/3}$ with the mean being of
order $N$, and that
the asymptotic distribution of appropriately scaled variables is
described by
the  Tracy--Widom distribution, which was
originally identified with the limit distribution 
for the largest eigenvalue of the Gaussian unitary
random matrix~\cite{TraWid94a}.
See Refs.~\citen{AldoDian99a,PDeif00a,BaikRain99c,KJohan02a} for
a  review.

In this paper motivated from results in
Ref.~\citen{BaikRain99b} and conjectures in Ref.~\citen{EMRain00a},
we introduce a physical  model of the vicious walkers based on the
hook Young tableaux.
We shall study the scaling limit of certain
probability, and clarify a relationship with the Tracy--Widom
distribution.

\begin{figure}[htbp]
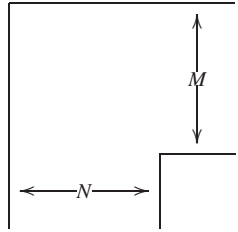

  \centering
  ~
  \xy <0.5cm,0cm>:
  (0,6) ; (6,6) **@{-} ,
  (0,0) ; (0,6) **@{-} ,
  (4,0) ; (4,2) **@{-} ,
  (4,2) ; (6,2) **@{-} ,
  \POS (0.3,1) \ar @{<->} (3.7,1) |N
  \POS (5,2.3) \ar @{<->} (5,5.7) |M
  \endxy
  
  \caption{($M,N$)-hook Young diagram must be contained in above
    ``hook'' region.}
  \label{fig:hook}
\end{figure}

For our later convention we define the ($M, N$)-hook Schur
functions
(or, sometimes called
the supersymmetric Schur function)~\cite{BerelRegev87a}
(see also Refs.~\citen{Macdo95,WFult97Book,Stanl99Book}), and
denote some properties
of
the hook Young tableaux briefly.
We set
$\mathbf{B}=\mathbf{B}_+  \sqcup \mathbf{B}_-$, and
\begin{align}
  \mathbf{B}_+
  &=
  \{ \epsilon_1, \dots, \epsilon_M \},
  &
  \mathbf{B}_-
  & =
  \{\epsilon_{M+1}, \dots, \epsilon_{M+N} \} .
  \label{define_B}
\end{align}
Hereafter we call $i$ as positive (resp. negative)
symbol when
$\epsilon_i\in\mathbf{B}_+$
(resp. $\epsilon_i\in\mathbf{B}_-$).
We fix an ordering in $\mathbf{B}$ as
\begin{equation}
  \label{define_order}
  \epsilon_1 \prec \epsilon_2 \prec \dots \prec \epsilon_{M+N}.
\end{equation}
It should be noted that, though we use
an ordering~\eqref{define_order},
following discussion can be applied for
any other choices of ordering
with $|\mathbf{B}_+|=M$ and $|\mathbf{B}_-|=N$.
For a given Young diagram $\lambda$, the
semi-standard  Young tableaux
(SSYT)
$T$ is given
by filling a number $1, 2, \dots, M+N$ in $\lambda$ by the
following rules;
\begin{itemize}
\item the entries in each row are increasing, allowing the repetition
  of
  positive symbols,
  but not permitting the repetition of
  negative symbols,

\item
   the entries in each column are increasing, allowing the repetition
   of
   negative symbols,
   but not permitting
   the repetition of
   positive symbols.
\end{itemize}
We define the
weight for SSYT $T$ as
\begin{equation}
  \wt(T)
  =\sum_{a=1}^{M+N} m_a \epsilon_a ,
\end{equation}
where $m_a$ is the number of $a$'s in $T$.
Then the hook Schur function $S_\lambda(x,y)$ is given by
\begin{equation}
  \label{define_hook}
  S_\lambda(x,y)
  =
  \sum_{
    \text{SSYT $T$ of shape $\lambda$}
  }
  \mathrm{e}^{\wt(T)} .
\end{equation}
Here we have used
\begin{equation*}
  \begin{cases}
    x_i = \mathrm{e}^{\epsilon_i},
    & \text{for $\epsilon_i \in \mathbf{B}_+$},
    \\[2mm]
    y_j = \mathrm{e}^{\epsilon_{M+j}},
    & \text{for $\epsilon_{M+j} \in \mathbf{B}_-$}.
  \end{cases}
\end{equation*}
The Schur function $s_\lambda(x)$
in usual sense corresponds to a case of
$\mathbf{B}_-=\emptyset$
($N=0$),
and 
the hook Schur function $S_\lambda$ with $\mathbf{B}_+=\emptyset$
($M=0$) reduces to the Schur
function for the conjugate partition $\lambda^\prime$;
\begin{align}
  S_\lambda(x,0) & = s_\lambda(x),
  &
  S_\lambda(0,y) & = s_{\lambda^\prime}(y) .
  \label{hook_and_Schur}
\end{align}
Due to the rule of filling a number, the Young diagram  $\lambda$
should be contained in the $(M,N)$-hook
(see Fig.~\ref{fig:hook}), and we have
\begin{equation*}
  S_\lambda(x,y)
  =
  \sum_{\mu \subset \lambda}
  s_\mu(x) \,
  s_{\lambda^\prime / \mu^\prime}(y) .
\end{equation*}
Furthermore when $\lambda$  contains the partition $(N^M)$, we have
\begin{equation*}
  S_\lambda(x,y)
  =
  s_\mu(x) \, s_\nu(y) \,
  \prod_{i=1}^M \prod_{j=1}^N (x_i + y_j) ,
\end{equation*}
where the partitions $\mu$ and $\nu$ are defined from $\lambda$
by
$\mu_i=\lambda_i - N$ and
$\nu_j = \lambda_j^\prime -M$, respectively.

The Jacobi--Trudi formula helps us to write the hook Schur function
as
\begin{equation}
  \label{Jacobi_Trudi}
  S_\lambda(x,y)
  =
  \det
  \left(
    c_{\lambda_i + j - i}
  \right)_{1 \leq i, j \leq \length(\lambda)} ,
\end{equation}
Here $\ell(\lambda)$ is the length of $\lambda$,
and $c_n$ is given by the generating function,
\begin{equation}
  H(t;x) \,
  E(t;y)
  =
  \sum_{n=0}^\infty c_n \, t^n ,
\end{equation}
with
\begin{align}
  H(t;x)
  & =
  \prod_j
  \frac{1}{
    1-t \,x_j
  },
  &
  E(t;y)
  & =
  \prod_j
  \bigl(
  1+t \, y_j
  \bigr) .
\end{align}

This paper is organized as follows.
In section~\ref{sec:walker}
we introduce a model of vicious walkers as a generalization of the
original model~\cite{MEFish84a}.
As far as we know, this model is first presented in this paper.
We define the bijection from path configurations of vicious walker to
the hook Young diagram.
Especially we show a relationship between the length of the Young
diagram and the number of movements of the first walker.
This type of bijection was first given in
Refs.~\citen{GuttOwczVien98a,PJForre01a} for the original vicious walker
model.
In section~\ref{sec:Toeplitz} we give the probability of
$\ell(\lambda)\leq \ell$ in terms of the Toeplitz determinant.
We further study the scaling limit of this probability based on the
transformation identity from the Toeplitz determinant to the Fredholm
determinant~\cite{BorodOkoun99a,BasoWido99b,ABottch00a}
 in section~\ref{sec:scaling}.
We apply the saddle point method to the Fredholm determinant following
Refs.~\citen{GravTracWido00a,GravTracWido02a}, and show that the
scaling limit coincides with the Tracy--Widom
distribution for the GUE~\cite{TraWid94a}.
In section~\ref{sec:example} we consider some simple examples as a
reduction of our
model.
Both the Meixner and the Krawtchouk ensembles can be regarded as a
reduction of our vicious walker model.
The last section is for conclusion and discussions.
We briefly comment on the random word related with the hook Young tableaux.

\section{Vicious Walker}
\label{sec:walker}

We define a   model of the random  walkers
which is related with the hook Schur
function~\eqref{define_hook}.
The model
is a generalization of one
introduced in Ref.~\citen{MEFish84a}, and
as will be clarified later
an algebraic property of the partition function is nothing but an
identity in Ref.~\citen{BaikRain99b}.

\begin{figure}[htbp]
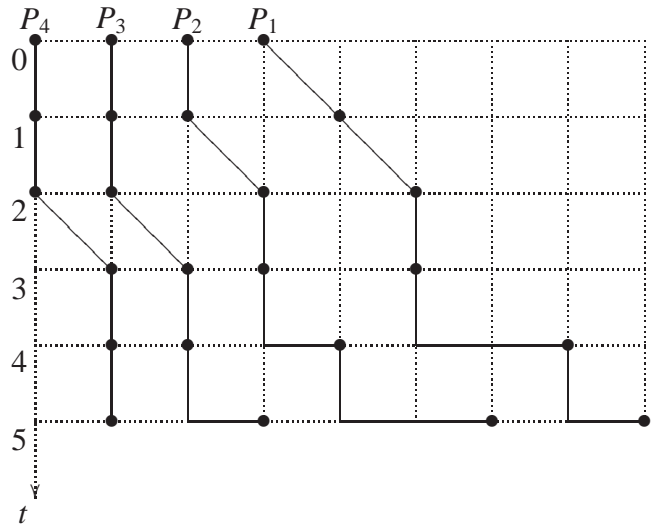

  \centering
  ~
  \xy 0;/r2.4pc/:
(0,0) ; (8,0) **@{.} ,
(0,1) ; (8,1) **@{.} ,
(0,2) ; (8,2) **@{.} ,
(0,3) ; (8,3) **@{.} ,
(0,4) ; (8,4) **@{.} ,
(0,5) ; (8,5) **@{.} ,
(0,5)  ; (0,-1) **@{.} ?> *\dir{>} *+!UR{t} ?<*+!UR{0} ?<(.17)*+!UR{1}
?<(.33)*+!UR{2} ?<(.5)*+!UR{3} ?<(.66)*+!UR{4} ?<(.83)*+!UR{5},
(1,0) ; (1,5) **@{.} ,
(2,0) ; (2,5) **@{.} ,
(3,0) ; (3,5) **@{.} ,
(4,0) ; (4,5) **@{.} ,
(5,0) ; (5,5) **@{.} ,
(6,0) ; (6,5) **@{.} ,
(7,0) ; (7,5) **@{.} ,
(8,0) ; (8,5) **@{.} ,
(3,5)="p11"; (4,4)="p12" **@{-} ?< *{\bullet}  *+!D{P_1} ?> *{\bullet},
"p12"; (5,3)="p13" **@{-}  ?> *{\bullet},
"p13"; (5,2)="p14" **@{-}  ?> *{\bullet},
"p14"; (5,1)="p15a" **@{-} ,
"p15a" ;(7,1)="p15" **@{-}  ?> *{\bullet},
"p15"; (7,0)="p16a" **@{-},
"p16a"; (8,0)="p16" **@{-}  ?> *{\bullet},
%
(2,5)="p21"; (2,4)="p22" **@{-} ?< *{\bullet}  *+!D{P_2} ?> *{\bullet},
"p22"; (3,3)="p23" **@{-}  ?> *{\bullet},
"p23"; (3,2)="p24" **@{-}  ?> *{\bullet},
"p24"; (3,1)="p25a" **@{-},
"p25a"; (4,1)="p25" **@{-}  ?> *{\bullet},
"p25"; (4,0)="p26a" **@{-} ,
"p26a"; (6,0)="p26" **@{-}  ?> *{\bullet},
%
(1,5)="p11"; (1,4)="p32" **@{-} ?< *{\bullet}  *+!D{P_3} ?> *{\bullet},
"p32"; (1,3)="p33" **@{-}  ?> *{\bullet},
"p33"; (2,2)="p34" **@{-}  ?> *{\bullet},
"p34"; (2,1)="p35" **@{-}  ?> *{\bullet},
"p35"; (2,0)="p36a" **@{-},
"p36a"; (3,0)="p36" **@{-}  ?> *{\bullet},
%
(0,5)="p41"; (0,4)="p42" **@{-} ?< *{\bullet}  *+!D{P_4} ?> *{\bullet},
"p42"; (0,3)="p43" **@{-}  ?> *{\bullet},
"p43"; (1,2)="p44" **@{-}  ?> *{\bullet},
"p44"; (1,1)="p45" **@{-}  ?> *{\bullet},
"p45"; (1,0)="p46" **@{-}  ?> *{\bullet},
\endxy

  \caption{Example of path configuration.
    For $t\leq 3$, a process is a ``normal'' evolution, while for
    $t \geq 4$
    it becomes a super time evolution.}
  \label{fig:exclusion}
\end{figure}

Evolution rule of vicious walkers is defined as follows.
Initially there are infinitely many walkers at
$\{ \dots, -2,-1 \}$, and we call each walker
$P_j$ whose initial
point is $-j$.
A walker is movable rightward if its right site is vacant.
Walkers $P_{j+1}, P_{j+2}, \dots$ are called successors of a walker
$P_j$ if they are next to each other in the order of the indices.
We consider two types of time evolution
(we assume that  there are totally $M+N$ time steps);
first $M$-steps are referred as ``normal'' time evolution, and
following $N$-steps are as ``super'' time evolution.
At a ``normal'' time evolution, a movable walker either stays  or moves
to its right together with an arbitrary number of its successors.
Thus we draw
\begin{align*}
  \text{move}
  &:
  \xy 0;/r1pc/:
  (0,-1)="ld" ;(2,-1)="rd" **@{.},
  "ld" ;(0,1)="lu" **@{.},
  "lu" ;(2,1)="ru" **@{.},
  "ru" ; "rd" **@{.},
  "lu" ; "rd" **@{-} ?> *{\bullet} ?< *{\bullet}
  \endxy
  &
  \text{stay}
  & :
  \xy 0;/r1pc/:
  (0,-1)="ld" ;(2,-1)="rd" **@{.},
  "ld" ;(0,1)="lu" **@{.},
  "lu" ;(2,1)="ru" **@{.},
  "ru" ; "rd" **@{.},
  "lu" ; "ld" **@{-} ?> *{\bullet} ?< *{\bullet}
  \endxy
\end{align*}
On the other hand,
at a ``super'' time evolution,
a  walker can move to its right
any number of  lattice units, though
$P_{j}$ cannot over-pass a position of $P_{j-1}$ at previous time.
To realize
this rule and to draw a non-intersecting path,
it is convenient to depict this step as
follows;
\begin{equation*}
  \underbrace{
    \xy 0;/r1pc/:
    (0,1)="s" ; (0,-1)="0" **@{-} ?< *{\bullet},
    (2,1) ; (2,-1) **@{.},
    (4,1) ; (4,-1) **@{.},
    (6,1) ; (6,-1) **@{.},
    (8,1)="1" ; (8,-1)="e" **@{.},
    "s" ; "1" **@{.},
    "0" ; "e" **@{-}   ?> *{\bullet}
    \endxy
  }_{
    \text{arbitrary number of lattice}
  }
\end{equation*}
Each path of the vicious walkers is required not to intersect each
other.
We see that the original model of vicious walkers~\cite{MEFish84a}
corresponds to
a case of $N=0$.
After $M+N$-time steps,
we denote $L_j(n)$ as the number of right moves made by the
walker $P_j$.
Here $n$ is a total number of movements of walkers.
In Fig.~\ref{fig:exclusion}
we give an example of path configuration of vicious walkers.
In this case we consider totally 5-time steps
($M=3$ and $N=2$), and
the total step of  right movements is $n=12$ with
$(L_1,L_2,L_3,L_4)
=(5,4,2,1)$.

It is now well  known for the model of the original vicious
walks~\cite{MEFish84a} that we have the bijection from the path
configuration
of  vicious walkers to
the Young diagram~\cite{GuttOwczVien98a}.
This bijection can be easily generalized to our model as follows.
For a path configuration
(see, \emph{e.g.}, Fig.~\ref{fig:exclusion}),
we draw Young tableaux $\lambda \vdash n$ with
$\lambda_j^\prime=L_j(n)$.
We insert in the $j$-th column from top
the 
times at which the $j$-th particle made a movement to its right.
Notice that, for a  super time evolution, we prepare
number of time as many as lattice units
the walker
moved.
For instance,
in a case of Fig.~\ref{fig:exclusion},
we put ``1 2 4 4 5'' in the first column, as $P_1$ moves $2$ lattice
units
rightward
at time-$4$.
Thus the top row is the list of times at which the walkers made their
first movement, the second row is the list of times at which the
walkers made their second movement, and so on.
It is clear
that the normal time corresponds to the positive symbol
$\mathbf{B}_+$ while the super time denotes the negative symbol
$\mathbf{B}_-$ in  SSYT.
The evolution rule 
supports a consistency  with ordering~\eqref{define_order}
in $\mathbf{B}$, and we know that the map is indeed the bijection.
Following this mapping
the path configuration in Fig.~\ref{fig:exclusion} is mapped to SSYT
given in Fig.~\ref{fig:young}.

\begin{figure}[htbp]
  \centering
  \young(1233,245,45,45,5)
  \caption{Semi-standard (hook) Young tableaux with entries from
    $\mathbf{B}_+= \{ 1, 2, 3  \}$ and
    $\mathbf{B}_-= \{ 4, 5  \}$.
  }
  \label{fig:young}
\end{figure}

To summarize,
we have a one-to-one correspondence between the path configuration and
SSYT;
when $n$ is the number of 
total moves of vicious walkers, we have $\lambda \vdash n$, and
the number $L_j( n)$ of right movements made by the $j$-th walker
is equal to the number of boxes in
the $j$-th column of SSYT.
Especially
the length $\ell(\lambda)$ of partition
coincides with $L_1(n)$.

\section{Partition Function and Toeplitz
  Determinant}
\label{sec:Toeplitz}

In the following we consider a model
where,
after totally  $n$-step movements  of the right movers,
every walkers return to their initial positions by totally
$n$-step left movements~\cite{PJForre01a}.
Here the number of 
normal (resp. super) time evolution  is supposed to be $M_1$
(resp. $N_1$) in the first right moves, while  the number of
normal (resp. super) time evolution is $M_2$ (resp. $N_2$) in the next
left moves returning to their initial positions.
The definition of normal and super time evolution in the left movers
simply follows from that of right movers as a mirror image.
Applying the bijection in previous section, the path configuration is
denoted
by pairs of SSYT $\lambda \vdash n$,
one is ($M_1,N_1$)-hook Young tableaux and another is ($M_2,N_2$)-hook
tableaux.

We denote $d_\lambda(M,N)$ as the number of semi-standard Young
tableaux of shape $\lambda$ with entries from
$\mathbf{B}_+ \sqcup \mathbf{B}_-$
(with $|\mathbf{B}_+|=M$ and
$|\mathbf{B}_-|=N$).
By definition, we have
$
S_\lambda(\underbrace{t,\dots,t}_M, \underbrace{t,\dots,t}_N)
= d_\lambda(M,N) \, t^n
$ for $\lambda \vdash n$,
and
once the Young diagram $\lambda$ is fixed
the number of SSYT $d_\lambda(M,N)$ corresponds to the number of path configuration with
fixed end points of right-moves.

We  have interests in  the probability  that
the number of right movements of the first walker $P_1$ is less
than $\ell$,
\begin{equation}
  \label{study_prob}
  \Prob(
  L_1  \leq \ell
  ) .
\end{equation}
Here the  probability ``Prob'' is defined  as follows;
we assign  the weight  $t$
(we set  $0<t<1$)
for every right- and left-moves,
and regard the weight of totally $n$-step walk as $t^n$.
Then each configuration of random walk, in which every walkers return
to their initial positions after total $2 \, n$-step,
is realized with
a probability
$t^{2n}/Z$.
An explicit form of 
the  normalization factor $Z$ will be  given later.
Based on the bijection map studied in previous section,
we find that the probability~\eqref{study_prob}
is given explicitly by
\begin{equation}
  \Prob(
  L_1 \leq \ell
  ) 
  =
  \frac{1}{Z} \,
  \sum_n
  \biggl(
  \sum_{
    \substack{\lambda \vdash n
      \\
      \ell(\lambda) \leq \ell
    }}
  d_\lambda(M_1,N_1) \,
  d_\lambda(M_2,N_2)
  \biggr) \,
  t^{2 n}
  .
\end{equation}
Note that
a normalization factor  is set to be
$\lim_{\ell \to \infty} \Prob(L_1 \leq \ell)=1$.

To relate this probability with the random matrix theory, we follow a
method in Ref.~\citen{TracWido99b}.
Applying the Gessel formula to
eq.~\eqref{Jacobi_Trudi},
we have
\begin{align}
  & \sum_{\ell(\lambda) \leq \ell}
  S_\lambda(x,y) \,
  S_\lambda(z,w)
  \nonumber \\
  & =
  \frac{1}{
    \ell ! \, (2 \, \pi)^\ell
  }
  \int_{- \pi}^{\pi}
  \mathrm{d} \theta \,
  \prod_{1\leq j < k\leq \ell}
  \left|
    \mathrm{e}^{\mathrm{i}  \theta_j}
    -
    \mathrm{e}^{\mathrm{i}  \theta_k}
  \right|^2
  \,
  \prod_{j=1}^\ell
  H(\mathrm{e}^{\mathrm{i} \theta_j} ; x) \,
  E(\mathrm{e}^{\mathrm{i} \theta_j} ; y) \,
  H(\mathrm{e}^{-\mathrm{i} \theta_j} ; z) \,
  E(\mathrm{e}^{-\mathrm{i} \theta_j} ; w) 
  \nonumber \\
  & =
  D_\ell(\varphi) ,
  \label{Gessel}
\end{align}
where $\varphi(z)$ is defined by
\begin{equation}
  \varphi(z)
  =
  \prod_{i,j}
  \frac{1+ y_i \, z^{-1}}{1-x_j \, z^{-1}}
  \,
  \frac{1+ w_i \, z}{1- z_j \, z} .
\end{equation}
We have used
$D_\ell(\varphi)$ as  the Toeplitz determinant for the function
$\varphi(z)$;
$D_\ell(\varphi)$ is the determinant of $\ell \times \ell$ matrix where
an $(i,j)$-element is given by
$\varphi_{i-j}$ with
$\varphi(z)=\sum_{n \in \mathbb{Z}} \varphi_n \, z^n$.
We note that
eq.~\eqref{Gessel} was also given in Ref.~\citen{BaikRain99b}.
Thus our model of random walkers corresponds to  a point process in
Ref.~\citen{BaikRain99b} which was introduced as a generalization of
Ref.~\citen{KJohan99a}.
We note that the strong Szeg\"o limit theorem gives a generalization
of the Cauchy formula,
\begin{equation}
  \lim_{\ell \to \infty}
  D_\ell( \varphi)
  =
  \prod_{i,j,m,n}
  \frac{
    (1+x_i \, w_n) \, ( 1 + y_j \, z_n)
  }{
    (1-y_j \, w_n) \, (1- x_i \, z_m)
  }.
\end{equation}

We now apply a principal specialization 
$\ps$ which set
$x_i=a \, q^i$ and $y_j=b \, q^j$~\cite{Stanl99Book}.
In  general we have
\begin{equation*}
  \ps (S_\lambda(x,y))
  =
  S_\lambda(
  \underbrace{a \, q , a \, q^2, \dots}_\infty,
  \underbrace{b \, q , b\, q^2 , \dots}_\infty
  )
  =
  q^{
    \sum_{i=1}^{\ell(\lambda)}
    i \, \lambda_i
  }
  \prod_{
    (i,j)\in \lambda
  }
  \frac{
    a+ b \, q^{j-i}
  }{
    1 - q^{\lambda_i -j + \lambda_j^\prime -i+1}
  } ,
\end{equation*}
and
for a case of $\lambda \vdash n$ and ($M,N$)-hook Young diagram,
by definition
we have by setting $a=b=t$ and $q=1$
\begin{equation*}
  \ps_{a=b=t;q=1} \Bigl(
  S_\lambda(x_1,\dots, x_M,
  y_1, \dots, y_N )
  \Bigr)
  =
  d_\lambda(M,N) \, t^n .
\end{equation*}
As a result, from eq.~\eqref{Gessel} we obtain the partition function as
\begin{equation}
  \label{general_sum}
  \sum_n
  \sum_{
    \substack{
      \ell(\lambda) \leq \ell
      \\
      \lambda \vdash n
      }}
    d_\lambda(M_1,N_1) \,
    d_\lambda(M_2,N_2) \,    
    t^{2 n}
    =
    D_\ell(\Tilde{\varphi}) ,
\end{equation}
where
\begin{equation}
  \label{kernel_walk}
  \Tilde{\varphi}(z)
  =
  \frac{(1+t \, z^{-1})^{N_1}}{(1- t \, z^{-1})^{M_1}}
  \,
  \frac{(1+t \, z)^{N_2} }{(1- t \, z)^{M_2}} .
\end{equation}
Due to the strong  Szeg\"o limit theorem, we obtain
a normalization factor $Z$ as
\begin{equation}
  Z=
  \lim_{\ell \to \infty}
  D_\ell(\Tilde{\varphi})
  =
  \frac{
    (1+t^2)^{M_1  N_2 + M_2 N_1}
  }{
    (1-t^2)^{M_1 M_2 +N_1 N_2}
  } .
  \label{define_Z}
\end{equation}
Combining these results, we get
\begin{equation}
  \Prob(
  L_1 \leq \ell
  )
  =
  \frac{1}{Z} \,
  D_\ell(
  \Tilde{\varphi}
  ).
\end{equation}

\section{Scaling Limit}
\label{sec:scaling}

We study the asymptotic behavior of the
probability~\eqref{study_prob}.
We note that
in Ref.~\citen{EMRain00a} the property of the scaling limit was
conjectured.
For our  purpose,
it is generally useful to rewrite the Toeplitz determinant
with the Fredholm determinant.
In fact,
once we know the Toeplitz determinant, it is possible to rewrite it in
terms of the Fredholm
determinant~\cite{BorodOkoun99a,BasoWido99b,ABottch00a}.
Namely we have
\begin{equation}
  \label{Fredholm}
  D_\ell(\Tilde{\varphi})
  =
  Z \,
  \det(1 - \mathcal{K}_\ell) ,
\end{equation}
where
$Z$ is defined in eq.~\eqref{define_Z}, and
$\mathcal{K}_\ell$ is the matrix defined by
\begin{equation}
  \label{Fredholm_element}
  \mathcal{K}_\ell(i,j)
  =
  \sum_{k=0}^\infty
  \bigl(
  \Tilde{\varphi}_- / \Tilde{\varphi}_+
  \bigr)_{i+\ell+k+1} \,
  \bigl(
  \Tilde{\varphi}_+ / \Tilde{\varphi}_-
  \bigr)_{-j-\ell-k-1}  .
\end{equation}
Here a subscript denotes the Fourier component of the function,
and we have used the Wiener--Hopf factor of $\Tilde{\varphi}$,
$\Tilde{\varphi}= \Tilde{\varphi}_+ \, \Tilde{\varphi}_-$,
\begin{align*}
  \Tilde{\varphi}_+
  & =
  \frac{(1+t \, z)^{N_2} }{(1- t \, z)^{M_2}} ,
  &
  \Tilde{\varphi}_-
  & =
  \frac{(1+t \, z^{-1})^{N_1}}{(1- t \, z^{-1})^{M_1}} .
\end{align*}
Note that we have set
$0<t<1$.
The probability~\eqref{study_prob} is now written by the Fredholm
determinant as
\begin{equation}
  \Prob(L_1 \leq \ell)
  =
  \det(1- \mathcal{K}_\ell) .
  \label{prob_Fredholm}
\end{equation}

Using a representation~\eqref{prob_Fredholm} in terms of the Fredholm
determinant, we study an asymptotic behavior by applying the saddle
point method following Refs.~\citen{GravTracWido00a,GravTracWido02a}.
We consider a limit $M_a , N_a \to \infty$ for $a=1,2$  with
fixed values;
\begin{align*}
  \frac{M_1}{N_2} & = m_1 ,
  &
  \frac{N_1}{N_2} & = n_1 ,
  &
  \frac{M_2}{N_2} & = m_2 .
\end{align*}
In the Fredholm determinant~\eqref{Fredholm_element}, matrix elements
are computed as
\begin{align*}
  (\Tilde{\varphi}_+/\Tilde{\varphi}_-)_{-\ell-j-k-1}
  & =
  \oint
  \frac{ \mathrm{d} \, z}{2 \, \pi \, \mathrm{i}} \,
  \frac{
    (1 + t\, z)^{N_2}
  }{
    (1-t \, z)^{M_2}
  }
  \cdot
  \frac{
    (1 - t/ z)^{M_1}
  }{
    (1+t / z)^{N_1}
  }
  \,
  z^{j+k+\ell} ,
  \\[2mm]
  (\Tilde{\varphi}_-/\Tilde{\varphi}_+)_{\ell+i+k+1}
  & =
  \oint
  \frac{ \mathrm{d} \, z }{2 \, \pi \, \mathrm{i}} \,
  \frac{
    (1-t \, z)^{M_2}
  }{
    (1 + t\, z)^{N_2}
  }
  \cdot
  \frac{
    (1+t / z)^{N_1}
  }{
    (1 - t/ z)^{M_1}
  }
  \,
  z^{-i-k-\ell-2} .
\end{align*}
A path of integration in the former integral
is chosen in a way that
it  surrounds $z=-t$, and
that  $z=1/t$ is outside.
On the other hand,
a  path of the latter integral  includes both
$z=0$ and $z=t$ while it excludes $z=-1/t$.
We set
\begin{equation}
  \label{ell_and_c}
  \ell = c \, N_2 + s \, N_2^{~\frac{1}{3}} ,
\end{equation}
where $c$ is  to be fixed later.
For brevity,
we define the function $\sigma(z)$ by
\begin{equation}
  \sigma(z)
  =
  m_1 \, \log (t-z)
  - n_1 \, \log(t+z)
  + \log(1+ t\, z) - m_2 \log(1-t \,z)
  +(-m_1+n_1+c) \, \log z .
\end{equation}
Then above integrals are given by
\begin{align*}
  (\Tilde{\varphi}_+/\Tilde{\varphi}_-)_{-\ell-j-k-1}
  & =
  (-1)^{M_1}
  \oint
  \frac{ \mathrm{d} \, z}{2 \, \pi \, \mathrm{i}} \,
  \mathrm{e}^{N_2 \, \sigma(z)} \,
  z^{j+k+s \, N_2^{~1/3}}
  \equiv
  (-1)^{M_1} \,
  I_1,
  \\[2mm]
  (\Tilde{\varphi}_-/\Tilde{\varphi}_+)_{\ell+i+k+1}
  & =
  (-1)^{M_1}
  \oint
  \frac{ \mathrm{d} \, z }{2 \, \pi \, \mathrm{i}} \,
  \mathrm{e}^{- N_2 \, \sigma(z)} \,
  z^{ -s \, N_2^{~1/3} - i - k -2}
  \equiv
  (-1)^{M_1} \,
  I_2 .
\end{align*}
We scale matrix indices as
$(i, j, k )
\to
( N_2^{~1/3} \, x,
 N_2^{~1/3} \, y,
 N_2^{~1/3} \, w)$,
and we consider to
apply the saddle point method to integrals,
\begin{align*}
  I_1
  & =
  \int\limits_{\mathcal{C}^{+}}
  \frac{ \mathrm{d} \, z}{2 \, \pi \, \mathrm{i}} \,
  \mathrm{e}^{N_2 \, \sigma(z)} \,
  z^{N_2^{~1/3} (w+y+s)},
  \\[2mm]
  I_2  &=
  \int\limits_{\mathcal{C}^{-}}
  \frac{ \mathrm{d} \, z}{2 \, \pi \, \mathrm{i}} \,
  \mathrm{e}^{- N_2 \, \sigma(z)} \,
  z^{- N_2^{~1/3} (w+x+s)-2} ,
\end{align*}
in a limit $N_2\to\infty$.
In these integrals, we fix
a parameter $c$ in eq.~\eqref{ell_and_c}
so that we have a \emph{double} saddle point,
namely
as a solution of a
set of equations,
\begin{align*}
  \frac{\mathrm{d}\, \sigma(z)}{\mathrm{d} z } 
  =
  \frac{\mathrm{d}^2 \, \sigma(z)}{\mathrm{d} z^2 }
  = 0 ,
\end{align*}
\emph{i.e.},
\begin{subequations}
\begin{gather}
  \label{d_of_sigma}
  \frac{m_1}{1-z/t}
  +
  \frac{1}{1+t \,z}
  =
  c-m_2+1
  +
  \frac{m_2}{1-t \, z}
  +
  \frac{n_1}{1+z/t} ,
  \\[2mm]
  \frac{m_1}{(t-z)^2}
  -
  \frac{n_1}{(t+z)^2}
  =
  -\frac{c-m_1+n_1}{z^2}
  +
  \frac{m_2 \, t^2}{(1-t \,z)^2}
  -
  \frac{t^2}{(1+t \,z)^2} .
\end{gather}
\end{subequations}
This set of equations is rewritten as
\begin{subequations}
\begin{equation}
  c =
  \frac{t }{t - z_0} \, m_1
  -\frac{t}{t + z_0} \, n_1
  -\frac{t \, z_0}{1- t \, z_0} \, m_2
  -\frac{t \, z_0}{1+ t \, z_0} ,
  \label{c_equation}
\end{equation}
where $z_0$ satisfies
\begin{equation}
  \label{z0_equation}
  \frac{m_1}{(t - z_0)^2}
  +
  \frac{n_1}{(t + z_0)^2}
  =
  \frac{1}{(1 + t \, z_0)^2}
  +
  \frac{m_2}{(1 - t \, z_0)^2}  .
\end{equation}
\end{subequations}
We see that
eq.~\eqref{z0_equation} always has a real solution in
$(-1/t, -t)$ as  far as $n_1 \neq 0$ because
$\text{l.h.s. $-$ r.h.s.}$ of eq.~\eqref{z0_equation} changes from
$-\infty$ to $\infty$ in $z\in ( -1/t, -t)$.
Generally real solutions of eq.~\eqref{z0_equation} are not only in
$(-1/t, -t)$, but
to deform paths of integrals adequately
we see that $z_0 \in ( -1/t, -t )$ is a
unique candidate of a
\emph{double} saddle point.
For example,
in a case of $m_1=m_2$ and $n_1=1$,
real solutions of  eq.~\eqref{z0_equation} are only $z=\pm 1$, and
we can conclude that a double saddle point
should be  $z_0=-1$ from a discussion below.
In  a case of
$m_1=n_1\geq 1$ and $m_2=1$, real solutions of eq.~\eqref{z0_equation}
are in
$(-1/t,-t)$, $(t,1/t)$,
$(-\infty, -1/t)$, and $(1/t,\infty)$
(the last 2 solutions exist only if
$m_1=n_1 \geq 1/t^2$), and from a
discussion to deform contours we see that only
$z_0\in (-1/t,-t)$ is 
possible as a double saddle point.
Based on these cases, it may be natural to conclude that
we choose
$z_0\in (-1/t,-t)$
as  a double saddle point.

\begin{figure}[htbp]
  \centering
  \begin{psfrags}
    \psfrag{A}{$\mathcal{C}^+$}
    \psfrag{B}{$\mathcal{C}^-$}
    \psfrag{Z}{$z_0$}
    \epsfig{file=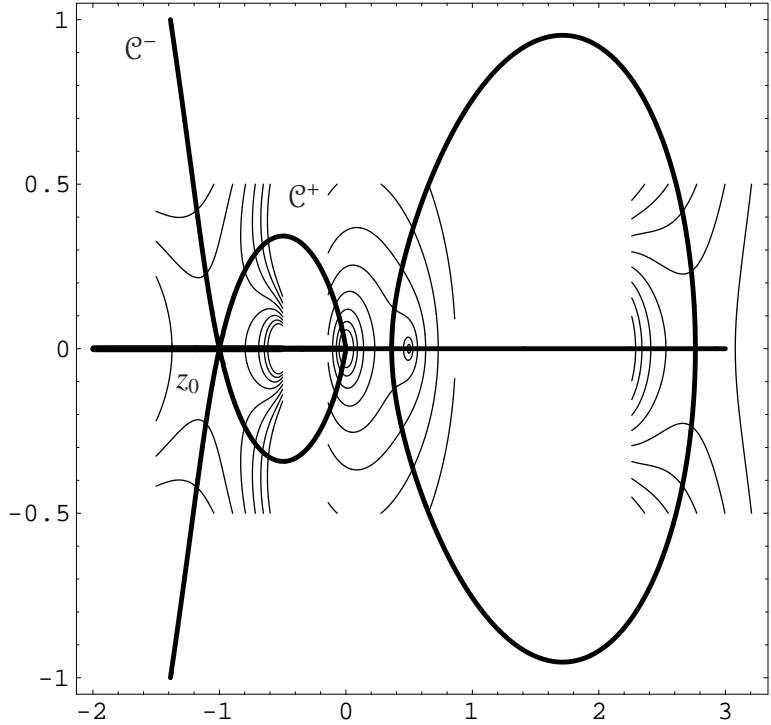}
  \end{psfrags}
  \caption{Typical example of
    the steepest descent path $\mathcal{C}^{\pm}$
    is depicted as a bold line.
    Here we have set $m_1=n_1=m_2=1$, and $t=1/2$.
    A double saddle point is $z_0=-1$, and other
    (simple) saddle points are
    $(25 \pm 3 \sqrt{41})/16$.
    We see that  paths $\mathcal{C}^{+}$
    and $\mathcal{C}^{-}$
    come to
    a double saddle point $z_0$ at angles
    $\pm \pi/3$ and $\pm 2 \, \pi/3$ respectively,
    and that
    another contour comes into
    (simple) saddle points
    with angle $\pm \pi/2$.
    A thin line denotes a local structure of
    the real part of the integrand around saddle points.
  }
  \label{fig:contour}
\end{figure}

Hereafter
we set a double saddle point $z_0$ 
so that $z_0 \in (-1/t, -t)$,
and fix a parameter $c$ by eq.~\eqref{c_equation}.
With  $z_0 \in (-1/t, -t)$,
we find  that $c>0$ from a definition~\eqref{c_equation}. 
With this choice of parameters, the fourth order
equation~\eqref{d_of_sigma} has a real
solution $z_0$ of multiplicity two, and other 2 solutions are in
$(-t, t)$
and
$(1/t,\infty)$.
Around $z_0$, we have a steepest descend path as in
Fig.~\ref{fig:contour}.
As $z_0$ is a double saddle point, paths come into $z_0$ with angles
$\pm \pi/3$ and $\pm 2\,\pi/3$.
Following Ref.~\citen{GravTracWido00a}, we denote such paths as
$\mathcal{C}^+$ and $\mathcal{C}^-$ respectively.
We see that
original paths  explained
above eq.~\eqref{ell_and_c}  can be deformed smoothly to contours
$\mathcal{C}^{\pm}$ avoiding their singularities.
Furthermore, we have
\begin{align*}
  &
  \frac{1}{2} \,
  \left.
    \frac{\mathrm{d}^3 \, \sigma(z)}
    {\mathrm{d} z^3}
  \right|_{z=z_0}
  \\
  & =
  \frac{t}{z_0^{~2}} \,
  \biggl(
  \frac{t-2 \, z_0}{(t-z_0)^3} \, m_1
  +
  \frac{t+2 \, z_0}{(t+z_0)^3} \, n_1
  -
  \frac{1- 2 \,t \, z_0}{(1-t \,z_0)^3} \, m_2
  -
  \frac{1+ 2 \,t \, z_0}{(1+t \,z_0)^3}
  \biggr)
  \\
  & =
  \frac{ -z_0 }{1-t \, z_0}
  \left(
    \frac{1-t^2}{(t-z_0)^3} \, m_1
    -
    \frac{1+t^2}{(t+z_0)^3} \, n_1
    +
    \frac{2 \, t}{(1+ t\, z_0)^3}
  \right) ,
%
\end{align*}
where in the first equality we have used eq.~\eqref{c_equation} to
delete a parameter $c$,
and in the second equality we have erased $m_2$ using
eq.~\eqref{z0_equation}.
Recalling $z_0 \in (-1/t, -t)$ and $0<t<1$, we see that
\begin{equation}
  \left.
    \frac{\mathrm{d}^3 \, \sigma(z)}
    {\mathrm{d} z^3}
  \right|_{z=z_0}
  >0,
  \label{3rd_order}
\end{equation}
which shows that
functions $|\mathrm{e}^{ \pm N_2 \, \sigma(z_0)}|$
have a maximum  value at $z=z_0$ on a contour $\mathcal{C}^{\pm}$.


With these settings, we have from the integral $I_1$ that
\begin{align*}
  & N_2^{1/3} \,
  \int\limits_{\mathcal{C}^+}
  \frac{\mathrm{d} \,  z}{2 \, \pi \, \mathrm{i}}
  \,
  \mathrm{e}^{N_2 \, \sigma(z)} \,
  z^{N_2^{~1/3} \, (w+y+s)}
  \\
  & =
  N_2^{1/3} \,
  \mathrm{e}^{N_2 \, \sigma(z_0)}
  \int\limits_{\mathcal{C}^+}
  \frac{\mathrm{d} \,  z}{2 \, \pi \, \mathrm{i}} \,
  \mathrm{e}^{\frac{N_2}{6} \, \sigma^{\prime \prime \prime}(z_0)
    \,
    (z-z_0)^3}
  \,
  z^{N_2^{~1/3} \, (w+y+s) }
  \\
  & =
  N_2^{1/3} \, z_0^{N_2^{~1/3} \,  (w+y+s)} \,
  \mathrm{e}^{N_2 \, \sigma(z_0)} 
  \int\limits_{\mathcal{C}^+}
  \frac{\mathrm{d} \, z}{2 \, \mathrm{i} \, \pi} \,
  \mathrm{e}^{
    \frac{N_2}{6} \, \sigma^{\prime\prime\prime}(z_0) \,
    z^3 +
    N_2^{~1/3} \, \frac{w+y+s}{z_0}\,  z
  }
  \\
  & =
  - z_0^{N_2^{~1/3} \, (w+y+s)} \,
  \mathrm{e}^{N_2 \, \sigma(z_0)} \,
  \frac{z_0}{\sigma} \,
  Ai(\frac{w+y+s}{\sigma}) .  
\end{align*}
Here $Ai(x)$ is the Airy function,
\begin{equation*}
  Ai(z) = \int_{-\infty}^\infty \frac{\mathrm{d} t}{2 \, \pi} \,
  \mathrm{e}^{\mathrm{i} \,
    (z \, t + t^3/3)
  },
\end{equation*}
and 
we have set a parameter $\sigma$ as
\begin{equation}
  \label{define_sigma}
  \sigma
  =
  - z_0 \,
  \left(
    \frac{1}{2}
    \left.
    \frac{\mathrm{d}^3 \, \sigma(z)}{\mathrm{d} z^3}
    \right|_{z=z_0}
  \right)^{1/3} .
\end{equation}
We have  $\sigma>0$ from eq.~\eqref{3rd_order}.
In the same way, we  have from the integral $I_2$ that
\begin{equation*}
  N_2^{~1/3}
  \int\limits_{\mathcal{C}^{-}}
  \frac{ \mathrm{d} \, z}{2 \, \pi \, \mathrm{i}} \,
  \mathrm{e}^{- N_2 \, \sigma(z)} \,
  z^{- N_2^{~1/3} (w+x+s)-2}
  =
  - z_0^{-N_2^{~1/3} \, (w+x+s)} \,
  \mathrm{e}^{- N_2 \, \sigma(z_0)} \,
  \frac{1}{z_0 \, \sigma} \,
  Ai(\frac{w+x+s}{\sigma}) .  
\end{equation*}
We then see that the kernel of the Fredholm
determinant~\eqref{Fredholm_element} is
given by the Airy kernel,
\begin{equation*}
  \frac{1}{\sigma^2}
  \int_0^\infty
  \mathrm{d} w \,
  Ai(\frac{s+x+w}{\sigma}) \,
  Ai(\frac{s+y+w}{\sigma})
  =
  \frac{1}{\sigma}
  \int_0^\infty
  \mathrm{d} w \,
  Ai(\frac{s+x}{\sigma}+w) \,
  Ai(\frac{s+y}{\sigma}+w) .
\end{equation*}

As a result, we obtain
\begin{equation}
  \lim_{N_2 \to \infty}
  \Prob
  \left(
    \frac{
      L_1
      -
      c \, N_2
    }{
      \sigma \, N_2^{~1/3}
    }
    \leq s
  \right)
  =
  F_2(s) .
  \label{scaling_TW}
\end{equation}
Here $F_2(s)$ is the 
Tracy--Widom distribution~\cite{TraWid94a} for
the scaling limit of the largest eigenvalue of the Gaussian unitary
ensemble, and is defined by
\begin{align}
  F_2(s)
  & =
  \det (1 - \mathcal{K}_{\text{Airy}})
  \\
  &=
  \exp
  \left(
    -
    \int_s^\infty \mathrm{d} x \, (x-s) \, q(x)^2
  \right) .
\end{align}
The second equality is from Ref.~\citen{TraWid94a}, and $q(x)$ is a
solution of the Painlev\'e II equation,
\begin{equation}
  q^{\prime\prime} = s \,  q + 2 \, q^3,
\end{equation}
with $q(s) \to Ai(s)$ in $s\to \infty$.

A proof of convergence would be done along a line of
Refs.~\citen{GravTracWido00a,GravTracWido02a}.

\begin{figure}[htbp]
  \centering
    \begin{psfrags}
    \psfrag{A}{$\mathcal{C}^+$}
    \psfrag{B}{$\mathcal{C}^-$}
    \psfrag{Z}{$z_0$}
    \epsfig{file=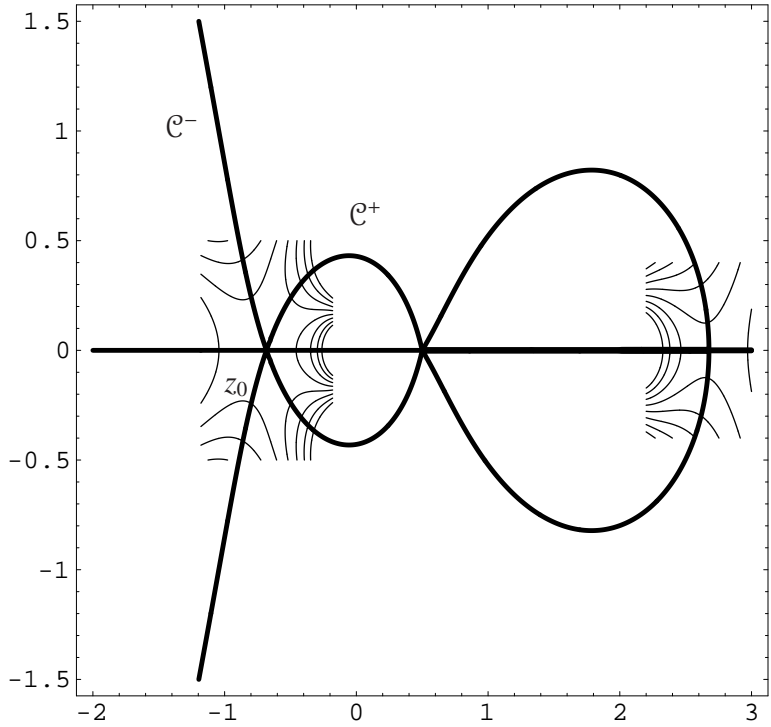}
  \end{psfrags}
  \caption{
    Example of the steepest descent path $\mathcal{C}^\pm$
    for a case of $n_1=0$ is depicted.
    Here we have set
    $m_1=4$, $m_2=1$, and $t=1/2$.
    A double saddle point is $z_0= -0.68254$, 
    and there is a (simple) saddle point at $2.67684$.
    As in Fig.~\ref{fig:contour},
    paths $\mathcal{C}^+$ and $\mathcal{C}^-$
    come to $z_0$  at angles
    $\pm \pi/3$ and $\pm 2 \, \pi/3$ respectively.
    Another contour comes into a (simple) saddle point with angle
    $\pm \pi/2$,  and  it ends at $t$.
    A thin line denotes a local structure of
    the real part of the integrand around saddle points.
  }
  \label{fig:contour_n_0}
\end{figure}

To close this section, we comment on a case of $n_1=0$.
In this case,
we further suppose  that $m_1 /t^2 >1+m_2$.
With this assumption, we see that there exists a real solution of
eq.~\eqref{z0_equation}
in $(-1/t, 0)$.
By setting this real solution to be
$z_0 \in (-1/t,0)$, we can prove
from eqs.~\eqref{c_equation} and~\eqref{define_sigma}
that $c>0$ and $\sigma>0$.
Note that with this setting of a parameter $c$ eq.~\eqref{d_of_sigma}
has a real solution $z_0$ of multiplicity two, and another solution is
in $(1/t, \infty)$.
See Fig.~\ref{fig:contour_n_0} as an example of a steepest descend
path.
As a result, we obtain the Tracy--Widom distribution~\eqref{scaling_TW}
as a scaling limit.

\section{Some Special Cases}
\label{sec:example}
\subsection{Meixner Ensemble}

We consider a case,
\begin{align*}
  & M_1=M_2=0 ,
  &
  &\text{\emph{i.e.}, $m_1=m_2=0$}.
\end{align*}
{}From the viewpoint of the random walkers,
the vicious walkers move obeying only super time
evolution rule.
In this case
the Toeplitz determinant~\eqref{general_sum} reduces to
\begin{align*}
  D_\ell
  \Bigl(
  (1+\frac{t}{z})^{N_1} \,
  (1 + t\,z)^{N_2}
  \Bigr)
  & =
  \sum_n
  \sum_{
    \substack{
      \ell(\lambda) \leq \ell
      \\
      \lambda \vdash n
    }}
  d_{\lambda^\prime} (N_1) \, d_{\lambda^\prime}(N_2) \,
  t^{2 n}
  \\
  & =
  \sum_n
  \sum_{
    \substack{
      \mu_1 \leq \ell
      \\
      \mu \vdash n
    }}
  d_{\mu} (N_1) \, d_{\mu}(N_2) \,
  t^{2 n} ,
\end{align*}
where $d_\lambda(N)$ denotes the number of (usual) semi-standard Young
tableaux, and we have
$d_\lambda(N)=d_\lambda(N,0)=d_{\lambda^\prime}(0,N)$
from eq.~\eqref{hook_and_Schur}.
The right hand side appeared in Ref.~\citen{KJohan99a}, and it gives
an example of the discrete orthogonal polynomial ensemble
as follows.
Using the hook formula~\cite{Stanl99Book},
\begin{equation}
  \label{hook_formula}
  d_\mu(M)
  =
  \prod_{1 \leq i< j \leq M}
  \frac{
    \mu_i - \mu_j + j -i
  }{
    j-i
  } ,
\end{equation}
the r.h.s. gives
\begin{equation*}
  \sum_n
  \sum_{
    \substack{
      \mu_1 \leq \ell
      \\
      \mu \vdash n
    }}
  \left(
    \prod_{1\leq i< j \leq N_2}
    \left[
      \frac{\mu_i - \mu_j + j -i}{j-i}
    \right]^2
    \cdot
    \left(
      \prod_{i=1}^{N_2}
      \left[
        \prod_{j=N_2-1}^{N_1}
        \frac{
          \mu_i + j-i
        }{j-i}
      \right]
      \,
      t^{2 \mu_i}
    \right)
  \right) ,
\end{equation*}
where we have assumed $N_1 \geq N_2$.
Introducing
\begin{equation}
  h_j = \mu_j + N_2 - j ,
\end{equation}
we obtain
\begin{multline}
  \Prob(L_1 \leq \ell)
  =
  (1-t^2)^{N_1 \, N_2}
  \, t^{-N_2 (N_2 -1)} \,
  \left[
    \prod_{j=0}^{N_2 -1}
    \frac{(N_1 - N_2)!}{j! \, (N_1 - N_2 + j)!}
  \right]
  \\
  \times
  \sum_{
    \substack{
      h \in \mathbb{N}^{N_2}
      \\
      \max\{h_i\} \leq \ell+N_2-1
    }}
  \left[
    \prod_{1 \leq i < j \leq N_2}
    (h_i - h_j)^2
  \right] \,
  \prod_{i=1}^{N_2}
  \begin{pmatrix}
    N_1 - N_2 + h_i \\
    h_i
  \end{pmatrix}
  \,
  t^{2 h_i} ,
\end{multline}
which is called the Meixner ensemble.

In fact using the Borodin--Okounkov identity~\eqref{Fredholm_element},
the kernel of the Fredholm determinant can be written in terms of the
Meixner polynomial
\begin{multline}
  (i-j) \, \mathcal{K}(i,j)
  =
  t^{4 N_2 + i+j} \, (1-t^2)^{N_1 -N_2 - 1} \,
  \begin{pmatrix}
    N_1+j \\
    N_2+j
  \end{pmatrix}
  \cdot
  \begin{pmatrix}
    N_1 \\
    N_2 
  \end{pmatrix}
  \cdot
  (-N_2)
  \\
  \times
  \biggl(
  M_{N_2}(i+N_2 ; N_1 - N_2 +1 ,t^2) \cdot
  M_{N_2 -1}(j+N_2 ; N_1 - N_2 + 1 , t^2)
  \\
  -
  M_{N_2 -1}(i+N_2 ; N_1 - N_2 +1 ,t^2) \cdot
  M_{N_2}(j+N_2 ; N_1 - N_2 + 1 , t^2)
  \biggr) ,
\end{multline}
which has a well known 
form of the correlation functions of the random
matrix (see, \emph{e.g.}, Ref.~\citen{Meh91}).
Note that the Meixner polynomial is defined by
\begin{equation*}
  M_n(x ; b, c)
  =
  {}_2 F_1
  \left(
    \begin{matrix}
      -n , -x \\
      b
    \end{matrix}
    ;
    1-\frac{1}{c}
  \right) .
\end{equation*}

A computation of the scaling limit can be done by the  method in
Section~\ref{sec:scaling}.
A double saddle point, $z_0 \in (-1/t , -t)$,
is explicitly solved as
\begin{equation*}
  z_0 = - \frac{t+\sqrt{n_1}}{1+t \, \sqrt{n_1}},
\end{equation*}
and we obtain
the Tracy--Widom
distribution~\eqref{scaling_TW} with parameters $c$ and $\sigma$
defined by
\begin{align}
  c & =
  \frac{
    t \, (2 \sqrt{n_1} + (n_1 + 1 ) \,t )
  }{
    1-t^2
  } ,
  \\[2mm]
  \sigma &=
  \frac{t^{1/3} \, (\sqrt{n_1} + t)^{2/3} \,
    (1+t \, \sqrt{n_1} )^{2/3}
  }{
    n_1^{~1/6} \, (1-t^2)
  } .
\end{align}
This result was derived by using the asymptotics of the Meixner
polynomial in
Ref.~\citen{KJohan99a}
(see also Ref.~\citen{BaDeMcMiZh01a}).

\subsection{Krawtchouk Ensemble}

We set
\begin{align}
  & N_1 = M_2 = 0 ,
  &
  &
  \text{
    \emph{i.e.},
    $n_1=m_2=0$
  },
\end{align}
The vicious walkers  obey a rule of normal time
evolution in the right moving, while they obey a rule of super time
evolution in the left moving.
In this case,  eq.~\eqref{general_sum} is read as
\begin{align*}
  D_\ell
  \left(
    \frac{(1+t \, z)^{N_2}}{(1-\frac{t}{z})^{M_1}}
  \right)
  =
  \sum_n
  \sum_{
    \substack{
      \ell(\lambda) \leq \ell
      \\
      \lambda \vdash n
    }}
  d_\lambda(M_1) \, d_{\lambda^\prime}(N_2)
  \, t^{2 n} .
\end{align*}
This becomes the Krawtchouk ensemble~\cite{KJohan99b} as follows
(this type of the Toeplitz determinant was also studied
in Ref.~\citen{GravTracWido00a}).
When
we substitute the hook formula~\eqref{hook_formula} into above expression,
we see that the r.h.s. reduces to
\begin{multline*}
  \left[
    \prod_{j=0}^{N_2 -1}
    \frac{(M_1+j)!}{  j!}
  \right]
  \sum_n
  \sum_{
    \substack{
      \mu_1 \leq \ell
      \\
      \mu \vdash n
    }}
  \left[
    \prod_{1 \leq i< j \leq N_2}
    (\mu_i - \mu_j + j-i)^2
  \right]
  \\
  \times
  \prod_{j=1}^{N_2}
  \frac{
    t^{2 \mu_j}
  }{
    (\mu_j + N_2 - j)! \,
    (M_1 + j - 1 - \mu_j)!
  } .
\end{multline*}
By use of
\begin{equation*}
  h_j = \mu_j + N_2 - j ,
\end{equation*}
this gives
\begin{multline}
  \Prob(L_1 \leq \ell)
  =
  (1+t^2)^{-M_1 \, N_2}
  \, t^{-N_2 (N_2 -1)} \,
  \left[
    \prod_{j=0}^{N_2 -1}
    \frac{(M_1 + j)!}{
      j! \, (N_2 + M_1 -1 )!
    }
  \right]
  \\
  \times
  \sum_{
    \substack{
      h \in \mathbb{N}^{N_2}
      \\
      \max\{h_i\} \leq \ell+N_2-1
    }}
    \left[
    \prod_{1 \leq i < j \leq N_2}
    (h_i - h_j)^2
  \right] \,
  \prod_{i=1}^{N_2}
  \begin{pmatrix}
    M_1 + N_2 -1 \\
    h_i
  \end{pmatrix}
  \,
  t^{2 h_i} ,
\end{multline}
which is the Krawtchouk ensemble.

The kernel of the
Fredholm determinant is  computed explicitly from
eq.~\eqref{Fredholm_element},
and it is given in terms of the Krawtchouk polynomial as a form of the
correlation functions;
\begin{multline}
  (i-j) \, \mathcal{K}(i,j)
  =
  - \frac{t^{i+j+4 N_2}}{(1+t^2)^{M_1 + N_2}} \,
  (M_1 + N_2 - 1) \,
  \begin{pmatrix}
    M_1 + N_2 -1 \\
    M_1 -1
  \end{pmatrix}
  \cdot
  \begin{pmatrix}
    M_1 + N_2 -1 \\
    N_2 + j
  \end{pmatrix}
  \\
  \times
  \biggl(
  K_{N_2} (i+N_2 ; \frac{t^2}{1+t^2} , M_1 + N_2 -1)
  \cdot
  K_{N_2 -1} (j+N_2 ; \frac{t^2}{1+t^2} , M_1 + N_2 -1)
  \\
  -
  K_{N_2-1} (i+N_2 ; \frac{t^2}{1+t^2} , M_1 + N_2 -1)
  \cdot
  K_{N_2} (j+N_2 ; \frac{t^2}{1+t^2} , M_1 + N_2 -1)
  \biggr) .
\end{multline}
Here the Krawtchouk polynomial is defined by
\begin{equation*}
  K_n(x ; p, N)
  =
  {}_2 F_1
  \left(
    \begin{matrix}
      -n , -x \\
      -N
    \end{matrix}
    ; \frac{1}{p}
  \right) .
\end{equation*}
We note that we have
\begin{equation*}
  K_n(x; p, N)
  =
  M_n( x ; -N , \frac{p}{p-1}) .
\end{equation*}

The scaling limit is also computed by the saddle point
method~\cite{GravTracWido00a}.
In this case we suppose $m_1 > t^2$, and
 we have a double saddle point $z_0\in(-1/t, 0)$ as
\begin{equation*}
  z_0= \frac{-\sqrt{m_1} + t}{1+t \, \sqrt{m_1} } ,
\end{equation*}
and we obtain the Tracy--Widom
distribution~\eqref{scaling_TW} where parameters $c$ and $\sigma$ are
defined from eqs.~\eqref{c_equation} and~\eqref{define_sigma} as
\begin{align}
  c &=
  \frac{
    t  \, (2 \sqrt{m_1} + (m_1 - 1) \, t)
  }{
    1+t^2
  } ,
  \\[2mm]
  \sigma
  & =
  \frac{
    t^{1/3} \, (\sqrt{m_1} - t)^{2/3} \,
    (1 + t \, \sqrt{m_1})^{2/3}
  }{
    m_1^{~1/6} \, (1+t^2)
  } .
\end{align}
One sees that this result
coincides with that of Ref.~\citen{KJohan99b} derived by 
use of asymptotics of the Krawtchouk polynomial.

\subsection{Symmetric Case}

We consider a case,
\begin{align}
  M_1&=M_2,
  &
  N_1&=N_2 ,
  &
  (i.e.,&
  &
  m_1 & = m_2=m,
  &
  n_1&=1
  ),
\end{align}
namely in  right- and left-movements we have equal number of
normal
and super time evolutions.
Unfortunately we are not sure whether this model is related with the
discrete orthogonal ensemble, but the parameters of the scaling
function can be simply solved as follows.

In a scaling limit $N_2\to\infty$, we obtain the Tracy--Widom
distribution~\eqref{scaling_TW} by applying the saddle point method.
In this case a double saddle point is $z_0=-1$, and
parameters in eq.~\eqref{scaling_TW} are computed from
eqs.~\eqref{c_equation} and~\eqref{define_sigma} as
\begin{align}
  \label{sym_c}
  c&=
  \frac{2 \, t \, \bigl( 1+m+(1-m) \,t \bigr)}{1-t^2},
  \\[2mm]
  \label{sym_sigma}
  \sigma
  &=
  \frac{
    t^{1/3} \, \bigl( m \, (1-t)^4 + (1+t)^4 \bigr)^{1/3}
  }{
    1-t^2
  } .
\end{align}

\section{Conclusion and Discussion}
\label{sec:conclusion}

We have introduced a generalization of the vicious walker model in
Ref.~\citen{MEFish84a}.
We find that there exists  a bijection map between the path
configuration of vicious walkers and the hook Young diagram as in the
case of the original vicious walkers.
We have exactly computed
a probability that the number of right movements of
the first walker is less than $\ell$, and have given a formula in
terms of the Toeplitz determinant.
We have further studied a scaling limit of the probability based on the
Borodin--Okounkov identity which relates the Toeplitz determinant with
the Fredholm determinant, and have  obtained
the Tracy--Widom
distribution for the largest eigenvalue of the Gaussian unitary random
matrix.
Other models which belong to
the orthogonal or the symplectic
universality classes  are for
future studies.

In the case of the vicious walker model,
crucial point is that there exists
the bijection map from the path configuration to a pair
of the semi-standard (hook) Young tableaux.
As was well studied~\cite{TracWido99b}, a pair of SSYT and the
standard tableaux is
related with the problem of the random word.
We can define the model of the random word which is related with the
hook Young diagram as follows~\cite{JFulm01b}.
We consider
a random word  by choosing from a set
$\mathbf{B}_+ \sqcup \mathbf{B}_-$ with
$\mathbf{B}_+=\{ 1, \dots, M\}$ and
$\mathbf{B}_-=\{M+1, \dots, M+N \}$.
When a word of length $n$ is given, 
we have
a generalization of the
Robinson--Schensted--Knuth (RSK)
correspondence~\cite{BerelRemme85a,KeroVers86a} (see also
Ref.~\citen{RegeSeem01a} for invariance under ordering of symbols);
we have a bijection between a word of length $n$ and pairs $(P,Q)$ of
tableaux of the same shape $\lambda \vdash n$
($P$ is SSYT from $\mathbf{B}$, and the recording tableaux $Q$ is the
standard Young
tableaux).
Rule to construct pairs of tableaux is essentially same with the
original RSK correspondence (see, \emph{e.g.},
Ref.~\citen{WFult97Book,Stanl99Book}), and
a difference is only that
negative symbols can bump himself while positive symbols cannot bump
himself.
Then for random word with length $n$,
the probability that
the length of longest decreasing
(strictly decreasing for positive symbols while weakly decreasing for
negative symbols)
subsequence 
is less
than or equal to
$\ell$ is then given by
\begin{equation}
  \sum_{
    \substack{
      \ell(\lambda) \leq \ell
      \\
      \lambda \vdash n
    }}
  d_\lambda(M,N) \, f^\lambda ,
  \label{prob_super_random}
\end{equation}
where $f^\lambda$ is the number of standard Young tableaux.

This can be rewritten in terms of the Toeplitz determinant based on
eq.~\eqref{Gessel}.
We use the exponential specialization~\cite{Stanl99Book},
\begin{equation}
  \ex(p_n) = t \, \delta_{1,n} ,
\end{equation}
where the power sum symmetric function $p_n$ is given by
\begin{equation*}
  p_n(x,y)
  =
  \sum_i
  x_i^{~n} +
  (-1)^{n-1} \sum_j y_j^{~n}
  .
\end{equation*}
Acting on the hook Schur function, we have
\begin{equation*}
  \ex \bigl(
  S_\lambda(x,y)
  \bigr)
  =
  f^\lambda \, \frac{t^n}{n !} ,
\end{equation*}
for $\lambda \vdash n$.
By applying the exponential specialization to
$(x,y)$ and the principal specialization  $\ps_{a=b=t;q=1}$ to
$(z,w)$ in
eq.~\eqref{Gessel},
we get
\begin{equation}
  \label{word_Fredholm}
  \sum_n
  \biggl(
  \sum_{
    \substack{
      \ell(\lambda) \leq \ell
      \\
      \lambda \vdash n
    }}
  d_\lambda(M,N) \, f^\lambda
  \biggr)
  \,\frac{t^n}{n !}
  =
  D_\ell(\Phi) ,
\end{equation}
where
\begin{equation}
  \label{kernel_Word}
  \Phi(z)
  =
  \mathrm{e}^{t/z} \,
  \frac{
    (1+z)^M
  }{
    (1-z)^N
  } .
\end{equation}
As a consequence,
the Poisson generating function of the
probability~\eqref{prob_super_random} is given by the Toeplitz
determinant of function $\Phi$.
As was seen from the fact that the kernel~\eqref{kernel_Word} can be
given from eq.~\eqref{kernel_walk} as an appropriate limit, the
scaling limit of eq.~\eqref{word_Fredholm}
reduces to the Tracy--Widom distribution as was shown in
Ref.~\citen{KJohan02a} for a case of $N=0$.
Detail will be discussed elsewhere.

It was shown in Ref.~\citen{TracWido99b}
that the generating  function~\eqref{word_Fredholm} with $N=0$ have an
integral representation in terms of solutions of Painlev\'e V
equation.
It remains for future studies to clarify a relationship between
the Toeplitz determinant~\eqref{word_Fredholm} in a case of $N\neq 0$ and
the Painlev\'e equations, especially integral solutions of the
Painlev\'e equation given in Ref.~\citen{ForreWitte02b}.

\bigskip
\noindent 
\underline{\textbf{Note Added:}}
After submitting  this paper,
Ref.~\citen{TracWido02c} appeared on net.
Therein
studied was  a limit theorem of
the  ``shifted Schur measure'',
where
the probability is defined  in terms of the Schur
$Q$-functions~\cite{Macdo95}.
To apply an a  method of Ref.~\citen{GravTracWido00a}
they obtained the Fredholm determinant after a finite perturbation of
a product of Hankel operator,
but their main result on a scaling limit  exactly coincides with
our results~\eqref{scaling_TW} with $M_1=N_1$ and $M_2=N_2$
(subsequently
one sees that their result for $\tau=1$
coincides with our above
results~\eqref{sym_c}--~\eqref{sym_sigma}
with $m=1$).
This coincidence may originate from a property of the Schur
$Q$-function.
The Schur $Q$-function is defined by filling ``marked'' and
``unmarked'' positive integers to the shifted Young diagram;
a rule of filling these numbers is much the same with a rule for the
semi-standard hook Young tableaux explained in Introduction,
once we identify  unmarked (resp. marked) numbers with
positive (resp. negative) symbols.
It will be interesting to investigate this connection in detail.

\newpage
\bibliographystyle{physics}


\begin{thebibliography}{10}

\bibitem{MEFish84a}
M.~E. Fisher,
\newblock J. Stat. Phys. {\bf 34}, 667 (1984).

\bibitem{GuttOwczVien98a}
A.~J. Guttmann, A.~L. Owczarek, and X.~G. Viennot,
\newblock J. Phys. A: Math. Gen. {\bf 31}, 8123 (1998).

\bibitem{PJForre01a}
P.~J. Forrester,
\newblock J. Phys. A: Math. Gen. {\bf 34}, L417 (2001).

\bibitem{JBaik00a}
J.~Baik,
\newblock Commun. Pure Appl. Math. {\bf 53}, 1385 (2000).

\bibitem{BaikDeifJoha99a}
J.~Baik, P.~Deift, and K.~Johansson,
\newblock J. Amer. Math. Soc. {\bf 12}, 1119 (1999).

\bibitem{TracWido99b}
C.~A. Tracy and H.~Widom,
\newblock Probab. Theory Relat. Fields {\bf 119}, 381 (2001).

\bibitem{KJohan99a}
K.~Johansson,
\newblock Commun. Math. Phys. {\bf 209}, 437 (2000).

\bibitem{Seppa01a}
T.~Sepp{\"a}l{\"a}inen,
\newblock Ann. Prob. {\bf 29}, 176 (2001).

\bibitem{PrahoSpohn00a}
M.~Pr{\"a}hofer and H.~Spohn,
\newblock Physica A {\bf 279}, 342 (2000).

\bibitem{GravTracWido00a}
J.~Gravner, C.~A. Tracy, and H.~Widom,
\newblock J. Stat. Phys. {\bf 102}, 1085 (2001).

\bibitem{Barys01a}
Y.~Baryshnikov,
\newblock Probab. Theory Relat. Fields {\bf 119}, 256 (2001).

\bibitem{TraWid94a}
C.~A. Tracy and H.~Widom,
\newblock Commun. Math. Phys. {\bf 159}, 151 (1994).

\bibitem{AldoDian99a}
D.~Aldous and P.~Diaconis,
\newblock Bull. Amer. Math. Soc. {\bf 36}, 413 (1999).

\bibitem{PDeif00a}
P.~Deift,
\newblock Notices Amer. Math. Soc. {\bf 47}, 631 (2000).

\bibitem{BaikRain99c}
J.~Baik and E.~M. Rains,
\newblock in \emph{Random Matrix Models and Their Applications}, edited by
  P.~Bleher and A.~Its,
  \emph{Mathematical Sciences Research
    Institute Publications}  \textbf{40},
  pp. 1--19, Cambridge Univ. Press, Cambridge, 2001.

\bibitem{KJohan02a}
K.~Johansson,
\newblock in \emph{European Congress of Mathematics, Vol. I},
edited by
C.~Casacuberta, R. Miro-Roig, J. Verdera, and S.~Xanbo-Descamps,
Prog. Math. \textbf{201},
pp. 445--456,
Birkh{\"a}user, 2001.

\bibitem{BaikRain99b}
J.~Baik and E.~M. Rains,
\newblock Duke Math. J. {\bf 109}, 1 (2001).

\bibitem{EMRain00a}
E.~M. Rains,
\newblock math.CO/0004082  (2000).

\bibitem{BerelRegev87a}
A.~Berele and A.~Regev,
\newblock Adv. Math. {\bf 64}, 118 (1987).

\bibitem{Macdo95}
I.~G. Macdonald,
\newblock \emph{Symmetric Functions and Hall Polynomials},
\newblock Oxford Univ. Press, Oxford, 2nd edition, 1995.

\bibitem{WFult97Book}
W.~Fulton,
\newblock \emph{Young Tableaux},
\newblock 
London Mathematical Society Student Texts \textbf{35},
Cambridge
  Univ. Press, Cambridge, 1997.

\bibitem{Stanl99Book}
R.~P. Stanley,
\newblock \emph{Enumerative Combinatorics. Vol.~2},
\newblock Cambridge Univ. Press, Cambridge, 1999.

\bibitem{BorodOkoun99a}
A.~Borodin and A.~Okounkov,
\newblock Integral Equations \& Operator Theory {\bf 37}, 386 (2000).

\bibitem{BasoWido99b}
E.~L. Basor and H.~Widom,
\newblock Integral Equations \& Operator Theory {\bf 37}, 397 (2000).

\bibitem{ABottch00a}
A.~B{\"o}ttcher,
\newblock Integral Equations \& Operator Theory {\bf 41}, 123 (2001).

\bibitem{GravTracWido02a}
J.~Gravner, C.~A. Tracy, and H.~Widom,
\newblock Ann. Prob. {\bf 30}, 1340 (2002).

\bibitem{Meh91}
M.~L. Mehta,
\newblock \emph{Random Matrices},
\newblock Academic Press, 1991.

\bibitem{BaDeMcMiZh01a}
J.~Baik, F.~Deift, K.~McLaughlin, P.~Miller, and X.~Zhou,
\newblock math.PR/0112162  (2001).

\bibitem{KJohan99b}
K.~Johansson,
\newblock Ann. Math. {\bf 153}, 259 (2001).

\bibitem{JFulm01b}
J.~Fulman,
\newblock math.CO/0104003  (2001).

\bibitem{BerelRemme85a}
A.~Berele and J.~Remmel,
\newblock J. Pure Appl. Alg. {\bf 35}, 225 (1985).

\bibitem{KeroVers86a}
S.~Kerov and A.~Vershik,
\newblock SIAM J. Alg. Disc. Method {\bf 7}, 116 (1986).

\bibitem{RegeSeem01a}
A.~Regev and T.~Seeman,
\newblock Adv. Appl. Math. {\bf 28}, 59 (2002).

\bibitem{ForreWitte02b}
P.~J. Forrester and N.~S. Witte,
\newblock math-ph/0204008  (2002).



\bibitem{TracWido02c}
C. A. Tracy and H. Widom,
\newblock math.PR/0210255 (2002).
\end{thebibliography}


\end{document}